\documentclass[onecolumn,noshowpacs,superscriptaddress,nobibnotes,nofootinbib,12pt]{revtex4}
\usepackage{amsmath,amssymb}
\usepackage{graphicx}

\begin{document}

\title{Deep inelastic and dipole scattering on finite length hot \\ $\mathcal{N}=4$ SYM matter\footnote{This work is supported in part by the US Department of Energy}}
\author{A. H. Mueller}
\email{amh@phys.columbia.edu}
\affiliation{Department of Physics, Columbia University, New York, NY, 10027, USA}

\author{A. I. Shoshi}
\email{shoshi@physik.uni-bielefeld.de}
\affiliation{Fakult{\"a}t f{\"u}r Physik, Universit{\"a}t Bielefeld, D-33501 Bielefeld,
Germany}

\author{Bo-Wen Xiao}
\email{bxiao@lbl.gov}
\affiliation{Nuclear Science Division, Lawrence Berkeley National Laboratory, Berkeley, CA 94720, USA}

\begin{abstract}
Deep inelastic scattering of $\mathcal{R}$-currents and the scattering of a small dipole on finite length hot $\mathcal{N}=4$ SYM matter are discussed. In each case we find the scale when scattering becomes strong is determined by a saturation momentum $Q^2_s \sim LT^3/x$ where $L$ is the length of the matter. For $\mathcal{R}$-currents we analyze the operator product expansion. For infinite length matter the series generated by the OPE is not Borel summable but we are able to determine the exponential part of the tunneling amplitude determining $F_2$ when $\frac{Q^2}{Q^2_s}\gg 1$  from the position of the singularity closest to the origin on the real axis of the Borel plane. In finite length matter the OPE series is not convergent but it is Borel summable. When a small dipole, and the string connecting the ends of the dipole, pass through hot matter there is an induced motion of the string in the $5^{th}$ dimension. When $T^4 L \cosh \eta$, with the $\eta$ the rapidity of the string, is large enough the string would normally break into several parts after leaving the medium, however, this cannot happen in the classical approximation in which we work.
\end{abstract}

\date{\today}
\maketitle
\newpage
\section{Introduction}
Our main object in this paper is to describe high energy deep inelastic scattering and the scattering of a small dipole on finite length hot matter in $\mathcal{N}=4$ SYM theory in the large
$N_c$ and large 't Hooft coupling limit. We suppose that the matter in question is static, that it is infinite extent in the $x$ and $y$-directions, but that it has length $L$ in the z-direction, the direction in which the current or dipole passes through the matter. As usual we take an $\mathcal{R}$-current with which to evaluate deep inelastic scattering\cite{Polchinski:2002jw, Hatta:2007he} and, when dipole scattering is considered, the ends of the dipoles are infinitely heavy "quarks" in the fundamental representation of $SU(N_c)$ so that the ends of the dipole are connected by a string\cite{Peeters:2006iu, Liu:2006he, Chernicoff:2006hi, Caceres:2006ta} which then passes through the matter.

In the calculations which are done it is always assumed that the coherence time of the projectile is much larger than the length $L$ which for $\mathcal{R}$-currents means that $\frac{2q}{Q^2}\gg L$ where $q$ and $Q^2$ are the momentum and virtuality of the current. We do not know an exact metric in $AdS_5 \times S^5$ space corresponding to the strip of hot matter, but we suppose that in the region corresponding to that outside the matter it is the normal $AdS_5 \times S^5$ metric while in the region corresponding to the matter that it is the three black brane metric with a short region of transition connecting these two domains. In the scattering of $\mathcal{R}$-currents one is clearly not sensitive to this transition region while in the scattering of the dipole(string) the situation is more subtle although, again, we believe there is little sensitivity to the details of the transition region.

All of our calculations are done in a classical supergravity approximation for $\mathcal{R}$-current scattering and a classical string approximation for dipole scattering. In the case of the $\mathcal{R}$-current calculation this should be a good approximation so long as $\frac{1}{x}\ll \sqrt{\lambda}$\cite{Polchinski:2002jw} where x is the Bjorken-$x$ variable and $\lambda$ is the 't Hooft coupling. Of course it would be very interesting to study the $\frac{1}{x}\gg \sqrt{\lambda}$  region where string excitations are important, but this is a formidable technical problem.

In the case of deep inelastic scattering in an infinite hot medium\cite{Hatta:2007cs, Hatta:2008tx, Iancu:2008sp} there is a sharp transition, say in $Q$, between a regime where $F_1$ and $F_2$ are large and a region where they are exponentially small. This transition region can be nicely characterized by the saturation momentum $Q_s \simeq T/x$ effective in the scattering. When $\frac{Q}{Q_s}\gg 1$ structure functions are very small while when $\frac{Q}{Q_s}\leq 1$  they are large. The gravitational wave corresponding to the $\mathcal{R}$-current does undergo multiple scattering when $\frac{Q}{Q_s}\gg 1$, corresponding to terms in the operator product expansion in the field theory description, however, in infinite extent matter these "gravitational" exchanges carry no momentum so that there is no production and $F_i$ remain zero except for a small tunneling contribution. When $\frac{T}{x}$ is as big as $Q$ the structure functions become nonzero in a nonperturbative way with respect to the multiple scattering peturbation series. In finite matter this picture changes. Now graviton exchange can transmit $z$-components of momentum due to the limited $z$-extent of the matter and there are small higher twist contributions to $F_i$ when $\frac{Q}{Q_s} \gg 1$ corresponding to specific terms in the multiple scattering series. We now briefly describe our main results.

In Sec.~\ref{sec2} we set up the formalism for calculating $F_i$ in finite length hot matter. Using an equation describing the time evolution of a gravity wave in the AdS-black hole metric, we show that over distances in the $z$-direction short compared to the coherence time of the wave an eikonal expression\cite{Levin:2008vj} gives an approximate solution. By simply replacing the single scattering term in the infinite matter formalism of Ref.~\cite{Hatta:2007cs} by the eikonal expression, over a length $L$, one has a formula for the forward scattering of the gravitational wave in finite matter. We find that scattering is strong when $Q^2\leq Q_s^2\sim \frac{LT^3}{x}$. Thus in finite matter the saturation momentum, the scale at which scattering becomes strong, now has the $x$-dependence expected from graviton exchange. This formula for $Q_s^2$, in agreement with that found in Ref.~\cite{Levin:2008vj} for cold matter, is in contrast to the formula for infinite matter where $Q_s^2 \sim \frac{T^2}{x^2}$ but the two formulas agree when one replaces $L$ in the finite matter formula by $L=\frac{1}{xT}$, the natural (coherence) length in the infinite medium. A major difference with the infinite matter case is that now the second order term in the multiple scattering series already contributes to the structure functions. The calculation of the double scattering contribution is given in the Appendix.~\ref{secondorder}, and that calculation should be reliable when $\frac{Q^2}{Q^2_s} \gg 1$.

In Sec.~\ref{sec3} we examine the form of the operator product expansion (multiple scattering series) for the retarded product of two $\mathcal{R}$-currents in the presence of the hot matter. We begin with infinite matter where the formalism has already been developed.\cite{Hatta:2007cs} The only available protected operator at small $x$ is the energy momentum tensor $T_{--}$ in terms of which the multiple scattering series is expressed. In contrast to the usual case in QCD, where the series is convergent for $x>1$ and where the divergence at $x=1$ signals production and nonzero $F_i$, we find the series is strongly divergent and that it is not Borel summable. We are able to evaluate the large order terms in the series, partly numerically, and, assuming the series is an asymptotic series given by a Borel representation, we determine the exponential part of $F_i$ in terms of the singularity on the positive real axis of the Borel plane nearest to the origin. Our result confirms the WKB tunneling caculation given in Ref.~\cite{Hatta:2007cs}. For finite matter the multiple scattering is divergent, but appears to be Borel summable.

In Sec.~\ref{sec4} we consider the scattering of a dipole on finite length hot matter which, on the gravity side of the AdS/CFT correspondence, corresponds to a string passing through a region, of length $L$, of an AdS-black hole metric. Before entering the black hole metric the string, whose ends are attached to the boundary of $AdS_5$-space, obeys equations of motion coming from the Nambu-Goto action in $AdS_5$-space. We do the calculation of the string passing through the region of the black-hole metric in two seemingly different ways which, however, give essentially the same picture of the string as it comes out of the black hole region. In our first procedure we expand the Nambu-Goto action about the $AdS_5$ metric and keep only the lowest order interaction with the black hole, the "one graviton" approximation. Using this lowest order term to calculate the equations of motion of the string and evaluate the new "initial conditions" for the evolution of the string in the $AdS_5$ metric. Upon leaving the black hole region the shape of the string is initially unchanged but it has velocity in the $5^{th}$-dimension as the main correction (see eq. (60)). We view this calculation as very close to the eikonal calculation which we did for $\mathcal{R}$-currents. The problem we are considering here is very close to that pioneered in Ref.~\cite{Albacete:2008ze}. The basic difference in our approaches is that we assume the string has not enough time to charge its shape as it passes through the matter while in Ref.~\cite{Albacete:2008ze} the assumption is that there is enough time for the string to relax to a static configuration while passing through the medium.

In our second procedure we treat the metric in the black hole region, corresponding to the strip of hot matter in the field theory side of the correspondence, more exactly. However, because of the sharp change in going from the $AdS_5$ to the $AdS_5$-black hole metric, the string must instantaneously slow down as it enters the black hole region. We evaluate this slowing down by using energy conservation. Then we use the, essentially exact, Nambu-Goto equations in the black hole metric to determine the motion of the string over the short interval during which it passes through the black hole region. While leaving the back hole region the string speeds up, in its motion along the $z$-direction. We find an identical answer for the induced motion of the string in the $5^{th}$-dimension due to passing through the black hole region as we found in our first procedure. Finally. we repeat the discussion in a heuristic way using a shock wave metric where the slowing down and speeding up described above are not visible, and we also do the calculation in the rest frame of the dipole in the Appendix.~\ref{restframe}.

Since we do not calculate quantum excitations on the string and we are not attempting to follow the (infinitely heavy) quark and antiquark at the ends of the string, there is a limited amount that we can calculate, in terms of cross sections, for the dipole going through the medium. When the saturation momentum of the medium gets as large as $u_m$, the minimum $u$-value of the string (see the metric (52)), we would normally expect strong radiation from the dipole. As we have seen earlier the saturation momentum for the medium is $Q_s^2 \sim T^3 L/x$. Setting $u_m=Q_s$ and $x=\frac{\gamma T}{u_m}$, with $\gamma$ the boost factor of the string, we expect, on the field theory side, radiation to occur when $\frac{Q_s^2}{u_m^2} >1$ or $\frac{T^4L\gamma}{u_m^3}>1$. When we carry out the string calculation we find that when the above equation is satisfied the lower part of the string, near $u_m$, lags behind the ends of the dipoles, after leaving the black hole regime, by an amount which would normally be too large for that part of the string to correspond to a part of the dipole. The lower part of the string could be expected to correspond to radiation. On the field theory side the picture is reasonably clear. The strip of matter is expected to free the parts of the fields having transverse momentum less than the saturation momentum\cite{Dominguez:2008vd}. At late times, the system would then consist of a freely moving dipole along with radiation which has cascaded to very low momentum. On the string theory side, this should correspond to a breaking of the string into at least two parts at large time. One part would correspond to a freely moving dipole and there must be another part, near $u=0$, corresponding to radiation. However, in the limit that we are working it does not appear possible for the string to split into two strings since that would correspond to a loop correction which is beyond the level at which we are working. From the string theory side of the correspondence there are strong oscillations on the string when $Q_s/u_m >1$. If we were quantizing the string this would correspond to diffractive excitations which would then ultimately decay at the level where even very small loop corrections were included. Thus while we believe that cross sections should become strong when $Q_s/u_m >1$, this picture can only be realized as strong radiation from the dipole when string loop corrections are included.

\section{Deep inelastic scattering on finite length hot matter}
\label{sec2}
In Ref.~\cite{Hatta:2007cs}, deep inelastic scattering on infinite extent hot $\mathcal{N}=4$ supersymmetric matter was studied using $\mathcal{R}$-currents instead of the usual electromagnetic currents in QCD. Here we extend that calculation to matter which has finite extent in the direction in which the current moves. For simplicity, we suppose the matter to be infinite in extent in the two directions orthogonal to the direction of motion of the current. More explicitly we suppose the current has momentum
\begin{equation}
q^{\mu} = \left( q^0 , q^1, q^2, q^3\right)=\left( \sqrt{q^2-Q^2} , 0, 0, q\right) \simeq\left( q-\frac{Q^2}{2q} , 0, 0, q\right)
\end{equation}
and the hot matter is confined to the region $0\leq z \leq L$, $-\infty < x, y < \infty$. We shall also limit our discussion to the case that $L$ is much less than coherence time of the current, that is
\begin{equation}
\frac{2q}{Q^2} \gg L.
\end{equation}
Using the same notation as in Ref.~\cite{Hatta:2007cs}, the structure functions are given by
\begin{equation}
F_1 = \frac{1}{2\pi} \textrm{Im} R_1 \label{f1d}
\end{equation}
and
\begin{equation}
F_2 = -\frac{n\cdot q}{2\pi T} \textrm{Im} R_2 \label{f2}
\end{equation}
where
\begin{equation}
R_{\mu \nu}=\left(\eta_{\mu \nu}-\frac{q_{\mu} q_{\nu}}{Q^2}\right) R_1+\left[n_{\mu } n_{\nu}- \frac{n\cdot q}{Q^2}\left(n_{\mu} q_{\nu}+n_{\nu} q_{\mu}\right)+\frac{q_{\mu} q_{\nu}}{\left(Q^2\right)^2}\left(n\cdot q\right)^2\right] R_2 \label{r12}
\end{equation}
and
\begin{equation}
R_{\mu \nu}\left(q\right) = i \int \textrm{d}^4 x \, e^{-iq\cdot x} \theta (x_0) \langle [J_\mu(x), J_\nu(0)]\rangle .\label{rmn}
\end{equation}
In Eqs.~(\ref{f2}) and (\ref{r12}), $n^{\mu}=\left(1, 0, 0, 0\right)$ and $\eta_{\mu \nu}=\left(-1, 1, 1, 1\right)$. The AdS/CFT correspondence allows one to evaluate $R_{\mu \nu}$ in terms of the classical action
\begin{equation}
S=-\frac{N^2T^2}{16}\int \textrm{d}^4 x
   \left[\left(A_0+\frac{\varpi}{k}A_3\right)a(u)
   -A_i\partial_uA_i(u)\right]_{u=0} =\int \textrm{d}^4 x \,\mathcal{S}\label{action1}
   \end{equation}
 as
\begin{equation}
R_{\mu\nu}(q)\,=\,\frac{\partial^2 \mathcal{S}}{\partial A_\mu
 \partial A_\nu}\,,
 \label{SR}
 \end{equation}
with $A_\mu = A_\mu(u=0)$ in Eq.~\ref{SR}. The gravitational field $A_{\mu}$ corresponds to the $\mathcal{R}$-current $J_{\mu}$ and in infinite matter obeys\cite{Son:2007vk}
\begin{eqnarray}
A_i^{\prime\prime}+\frac{f^\prime}{f}A_i^\prime +
   \frac{\varpi^2-k^2f}{uf^2}A_i\,&=&\,0 \label{ai} \\
a^{\prime\prime}+\,\frac{(uf)^\prime}{uf}\,a^\prime +
\,\frac{\varpi^2-k^2f}{uf^2}\, a\,&=&\,0 \label{a0}
\end{eqnarray}
with $a=A_{0}^{\prime}$ and primes in Eqs.~(\ref{ai}) and (\ref{a0}) denote $u$-derivatives with the metric
\begin{equation}
\textrm{d} s^2=\frac{(\pi TR)^2}{u}(-f(u)\textrm{d} t^2+\textrm{d}
 \vec{x}^2)+\frac{R^2}{4u^2f(u)}\textrm{d} u^2 +R^2 \textrm{d} \Omega^2_5\label{metric1}
\end{equation}
and where $A_{\mu}$  is a plane wave
\begin{equation}
A_{\mu}\left(t, \vec{x}, u \right) = e^{iqz-i\omega t} \, A_{\mu} \left(u\right).
\end{equation}
We are also using scaled variables
\begin{equation}
k=\frac{q}{2\pi T} \quad , \quad \varpi =\frac{\omega}{2\pi T},
\end{equation}
and $f\left(u\right)$ in Eq.~(\ref{metric1}) is given by $f\left(u\right)=1-u^2$.

From the gravity side of the correspondence, one can calculate $F_1$ and $F_2$ by evaluating the forward scattering amplitude for the gravitational waves $A_{i}$ and $a$. In order to do this we recall that the single scattering term was evaluated explicitly for infinite extent matter in Ref.~\cite{Hatta:2007cs}. The result for $a\left(\zeta\right)$ was
\begin{equation}
 a^{(1)}\left(\zeta\right) = \int_{0}^{\infty} \textrm{d} \zeta^{\prime} G\left(\zeta, \zeta^{\prime}\right) \frac{-k^2 \zeta^{\prime 4}}{16 K^6} a^{(0)}\left(\zeta^{\prime}\right)\label{a1}
\end{equation}
where
\begin{equation}
a^{(0)}\left(\zeta\right)=-2k^2 A_{L}\left(0\right) K_{0}\left(\zeta\right) \label{a0d}
\end{equation}
with $\zeta = 2 K\sqrt{u}$ and
\begin{equation}
A_{L}\left(0\right)=\left[A_{0}\left(u\right)+\frac{\varpi}{k}A_{3}\left(u\right)\right]_{u=0} \label{al}
\end{equation}
The Green's function is given by
\begin{equation}
G\left(\zeta, \zeta^{\prime}\right)=-\zeta^{\prime}\left[\textrm{K}_{0}\left(\zeta\right) \textrm{I}_{0}\left(\zeta^{\prime}\right)\Theta \left(\zeta-\zeta^{\prime}\right)+\textrm{K}_{0}\left(\zeta^{\prime}\right) \textrm{I}_{0}\left(\zeta\right) \Theta \left(\zeta^{\prime}-\zeta\right)\right].\label{green1}
\end{equation}
In the case of finite matter all that can change in this single scattering contribution is the normalization. In infinite extent matter the normalization of the single scattering term was set by
\begin{equation}
\langle T_{--} \rangle=\frac{4}{3} \langle T_{00} \rangle = \frac{\pi ^2 N^2 T^4}{4}
\end{equation}
after using the operator product expansion in Eq.~(\ref{rmn}). The amplitudes $R_1$ and $R_2$ have dimensions of an inverse area for infinite extent matter. For finite length matter what naturally occurs in the single scattering is
\begin{equation}
\int_{0}^{L} \textrm{d} z\langle T_{--} \rangle=\langle T_{--} \rangle L,
\end{equation}
however, in order to keep the dimension of $R_i$ and $F_i$, the same as for infinite mattter we will multiply by the temperature $T$ so that now our normalization of $a^{(1)}$ is given by Eq.~(\ref{a1}) along with an additional factor of $LT$ on the right hand side of that equation.

In order to get the higher multiple scatterings we use the time-dependent evolution equation for $\psi \left(\tilde{t},\chi\right)=\sqrt{\chi} \, a$ given by \cite{Hatta:2008tx}
\begin{equation}
i \frac{\partial \psi}{\partial \tilde{t}} = \left(-\frac{1}{2k}\frac{\partial^2}{\partial \chi^2}-\frac{1}{8k\chi^2}+\frac{K^2}{2k}-\frac{k \chi^4}{32}\right) \psi \label{psi1}
\end{equation}
where $\tilde{t}=2\pi T t$ and $\chi = 2\sqrt{u}=\zeta/K$. Eq.~(\ref{psi1}) is used in the region $0<z< L$ and we identify $2\pi T z =\tilde{z}$ with $\tilde{t}$. It is straightforward to check that
\begin{equation}
\psi \left(\tilde{z}, \chi\right) = \psi^{(0)}\left(\chi \right) \exp\left[\frac{ik\chi^4\tilde{z}}{32}\right] \label{so1}
\end{equation}
is an approximate solution to Eq.~(\ref{psi1}) so long as $k\chi /K \gg \tilde{L}$. Since the important region of $\chi$ is $\chi \sim 1/K$, this condition becomes $k/K^2\gg \tilde{L}$, the condition that the coherence time of the gravitational wave be much larger than the length of the matter.

Now it is a simple matter to insert the full eikonal solution Eq.~(\ref{so1}) in Eq.~(\ref{a1}) instead of the first order term in the potential to get
\begin{equation}
 a\left(\zeta\right) = \int_{0}^{\infty} \textrm{d} \zeta^{\prime} G\left(\zeta, \zeta^{\prime}\right)\left\{ \frac{1-\exp\left[i \frac{k \zeta^{\prime 4}\tilde{L}}{32K^4}\right]}{2\pi i} \frac{2k}{K^2}\right\} a^{(0)}\left(\zeta^{\prime}\right)+a^{(0)}\left(\zeta\right)\label{a1m}
\end{equation}
where the normalization of Eq.~(\ref{a1m}) is fixed by the agreement between Eq.~(\ref{a1m}) and Eq.~(\ref{a1}) when the single scattering term in $\left\{ \quad \right\}$ is taken and when Eq.~(\ref{a1}) is multiplied by $LT$ as discussed above.

Similarly for transverse fields one gets
\begin{equation}
 h\left(\zeta\right) = \int_{0}^{\infty} \textrm{d} \zeta^{\prime} G_{T}\left(\zeta, \zeta^{\prime}\right)\left\{ \frac{1-\exp\left[i \frac{k \zeta^{\prime 4}\tilde{L}}{32K^4}\right]}{2\pi i} \frac{2k}{K^2}\right\} h^{(0)}\left(\zeta^{\prime}\right)+h^{(0)}\left(\zeta\right)\label{h1m}
\end{equation}
where
\begin{equation}
A_{i} \left(\zeta\right)=\zeta h\left(\zeta\right) A_{i} \left(0\right) \label{ai2}
\end{equation}
and
\begin{equation}
h^{(0)}\left(\zeta\right) =\textrm{K}_{1} \left(\zeta\right)
\end{equation}
with
\begin{equation}
G_{T}\left(\zeta, \zeta^{\prime}\right)=-\zeta^{\prime}\left[\textrm{K}_{1}\left(\zeta\right) \textrm{I}_{1}\left(\zeta^{\prime}\right)\Theta \left(\zeta-\zeta^{\prime}\right)+\textrm{K}_{1}\left(\zeta^{\prime}\right) \textrm{I}_{1}\left(\zeta\right) \Theta \left(\zeta^{\prime}-\zeta\right)\right]. \label{gt}
\end{equation}

Using Eqs.~(\ref{a1m}) and (\ref{h1m}) in Eq.~(\ref{action1}) one finds \cite{Levin:2008vj}
\begin{equation}
\mathcal{S}-\mathcal{S}_0=\frac{iN^2}{32 \pi^2 x}\int_{0}^{\infty} \textrm{d}\zeta
\left\{ 1-\exp\left[i \frac{k \zeta^{ 4}\tilde{L}}{32K^4}\right]\right\}
   \left[Q^2 A_{i}^2\left(0\right)\zeta \textrm{K}_{1}^2\left(\zeta\right)+q^2 A_{L}^2\left(0\right)\zeta \textrm{K}_{0}^2\left(\zeta\right) \right]\label{action2}
   \end{equation}
 with $\frac{1}{x}=\frac{2qT}{Q^2}$ and where $\mathcal{S}_0$ is the action when $L=0$. When
 \begin{equation}
 Q^2\ll Q^2_s\sim LT^3/x \label{sam}
 \end{equation}
 the exponential factor in Eq.~(\ref{action2}) is zero and one gets
 \begin{equation}
\mathcal{S}=\frac{iN^2}{32 \pi^2 x}
   \left[\frac{1}{2}\left(q A_{0}+\omega A_3\right)^2+Q^2 A_{i}^2\ln \left(\frac{k\tilde{L}}{K^4}\right)^{1/4}\right]_{u=0}\label{action3}
   \end{equation}
leading to
\begin{eqnarray}
F_{1}&=&\frac{N^2 Q^2}{128\pi^3 x}\ln\frac{Q^2_s}{Q^2} , \label{f1f}\\
F_{L}&=& F_2-2xF_1=\frac{N^2Q^2}{32\pi^3} \label{f2f}
\end{eqnarray}

A number of comments may be useful here, (i) The $x$-dependence of the saturation momentum given in Eq.~(\ref{sam}) is the same as that found in Ref.~\cite{Hatta:2007he} for deep inelastic scattering on a dilaton. (ii) Eq.~(\ref{f1f}) has a natural parton interpretation. The number density of partons in the target, the slab of hot matter is, when $Q^2\ll Q^2_s$,
\begin{equation}
\frac{\textrm{d}N}{\textrm{d}^2 x_{\perp}}\sim F_2\sim N^2Q^2
\end{equation}
up to the logarithm in Eq.~(\ref{f1f}) which will be discussed next. This means that in the saturation region occupation numbers are, parametrically, on the order of one. (iii) The logarithm in Eq.~(\ref{f1f}) reflects the fact that before the quanta produced by the $\mathcal{R}$-current reach the target they have a logarithmically large number of components, when $Q^2_s/Q^2\gg 1$, which will be completely absorbed by the target. An identical logarithm exists in deep inelastic scattering in perturbative QCD, and with $N_c^2\rightarrow N_c$ (reflecting the number of quarks in QCD) Eq.~(\ref{f1f}) remains parametrically true in QCD.

The major difference between Eq.~(\ref{action2}) and the infinite matter formula of Ref.~\cite{Hatta:2007cs} is that Eq.~(\ref{action2}) gives nonzero, although higher twist, structure functions when $Q^2/Q_s^2\gg 1$ with the dominant contribution coming from the second order term in the expansion of the exponential. This is exactly as described for deep inelastic scattering on a dilaton in Ref.~\cite{Hatta:2007he}. Such contributions do not exist in infinite matter because the multiple graviton exchange, corresponding to the exponential in Eq.~(\ref{action2}), all have zero momentum transfer in infinite matter. In finite matter the gravitons can have nonzero momentum and can convert the space-like incoming gravitational wave to a time-like wave which then will decay giving nonzero $F_1$ and $F_2$. Eq.~(\ref{action2}) is a classical formula . There are no string excitations included. If $\frac{q}{L} \ll \frac{\sqrt{\lambda}}{R^2}$, string excitations can not be produced and Eq.~(\ref{action2}) should be quite accurate\cite{Hatta:2007he}. If $\frac{q}{L} \geq \frac{\sqrt{\lambda}}{R^2}$, the lack of inclusion of string excitations can not be justified, although we believe Eq.~(\ref{action2}) will be an approximate formula.

\section{The operator product expansion and multiple scattering}
\label{sec3}
In Ref.~\cite{Hatta:2007cs}, the multiple scattering series for scattering of the gravitational wave $A_{\mu}$ was discussed for infinite matter while the expansion of the exponential in Eq.~(\ref{action2}) gives the multiple scattering series for finite matter. On the field theory side of the AdS/CFT correspondence this series represents the operator product expansion of the $\mathcal{R}$-currents in Eq.~(\ref{rmn}) in terms of multiple powers of the only protected operator relevant at high energy\cite{Polchinski:2002jw}, the energy momentum tensor $T_{--}$. In QCD the proton matrix element of the retarded product of electromagnetic currents, analogous to the thermal expectation given in Eq.~(\ref{rmn}), has a convergent operator product expansion when $\frac{1}{x}=\omega_{bj}<1$ with the singularity at $\omega_{bj}=1$ corresponding to particle production and nonzero structure functions given by the imaginary part of $R$ when $\omega_{bj}>1$. In Ref.~\cite{Hatta:2007cs}, it was found that structure functions are exponentially small when $Qx/T \gg 1$. Does this mean that the operator expansion is convergent for $\omega_{bj}<Q/T$ with a singularity when $\omega_{bj}$ is of the order of $Q/T$? As we shall now see this is not the case. The operator product expansion, both for finite and for infinite matter, gives a divergent series for all values of $x$. For infinite matter this series is not Borel summable, however, the position of the singularity closest to the origin in the Borel plane can be used to evaluate the exponentially small tunneling amplitude giving $F_1$ and $F_2$ when $Qx/T \simeq 2\pi^2 K^3/k \gg 1$, which amplitude was first evaluated in Ref.~\cite{Hatta:2007cs} using WKB methods. For finite matter the operator product series is also divergent but it is Borel summable.

\subsection{Infinite extent matter}
Our starting point is the integral equation giving the operator product expansion - multiple scattering series, say for the field $a\left(\zeta\right)$
\begin{equation}
 a\left(\zeta\right) = a^{(0)}\left(\zeta^{\prime}\right)+\int_{0}^{\infty} \textrm{d} \zeta^{\prime} G\left(\zeta, \zeta^{\prime}\right) \frac{-k^2 \zeta^{\prime 4}}{16 K^6} a\left(\zeta^{\prime}\right)\label{aall}
\end{equation}
with $G\left(\zeta, \zeta^{\prime}\right) $ given in Eq.~(\ref{green1}) and $a^{(0)}$ in Eq.~(\ref{a0d}). The multiple scattering series is obtained by iteration of Eq.~(\ref{aall}). If one writes
\begin{equation}
a\left(\zeta , \rho \right)=-2\pi k^2 A_{L}\left(0\right)\sum_{N=0}^{\infty} a_{N}\left(\zeta\right) \rho^{N}, \label{series}
\end{equation}
where
\begin{equation}
\rho=\frac{k^2}{32K^6},
\end{equation}
then
\begin{equation}
a_{N}\left(\zeta\right)=\frac{(-2)^N}{\pi}\int \prod_{i=1}^{N} \zeta_{i}^4 \textrm{d}\zeta_i G\left(\zeta, \zeta_1\right)G\left(\zeta_1, \zeta_2\right)\cdots G\left(\zeta_{N-1}, \zeta_N\right) \mathrm{K}_{0}\left(\zeta_{N}\right).\label{an}
\end{equation}
Now for large $N$ all values of $\zeta_{i}$ in Eq.~(\ref{an}) will be large except $\zeta$ which we keep small in order to use Eq.~(\ref{an}) in Eq.~(\ref{action1}). Using
\begin{equation}
G\left(\zeta_{i}, \zeta_{i+1}\right)\simeq -\frac{1}{2}\sqrt{\frac{\zeta_{i+1}}{\zeta_{i}}} e^{-\left|\zeta_{i}-\zeta_{i+1}\right|}
\end{equation}
and
\begin{equation}
G\left(\zeta, \zeta_{1}\right)\simeq -\sqrt{\frac{\pi \zeta_{1}}{2}} e^{-\zeta_1}
\end{equation}
when $\zeta_{1}$, $\zeta_{i}$, $\zeta_{i+1}$ are large and $\zeta$ is small, one gets
\begin{equation}
a_{N}\left(0\right) \simeq \int \prod_{i=1}^{N} \zeta_{i}^4 \textrm{d}\zeta_i \exp \left[-\zeta_1-\left|\zeta_{1}-\zeta_{2}\right|-\left|\zeta_{2}-\zeta_{3}\right|\cdots -\left|\zeta_{N-1}-\zeta_{N}\right|-\zeta_{N}\right] \label{an0}
\end{equation}
when $N$ is large. We have been unable to evaluate  Eq.~(\ref{an0}) analytically, however, we have evaluated $a_{N}\left(0\right)=a_N$ numerically up to values of $N$ as large as 13. As shown in table \ref{nume} we find for large $N$ that
\begin{equation}
\frac{a_{N+1}}{a_{N}}=32\frac{\Gamma(4N+4)}{\Gamma(4N)}\left(\frac{3\sqrt{2\pi}}{2\Gamma^2\left(\frac{1}{4}\right)}\right)^4=\frac{32}{c^4}\frac{\Gamma(4N+4)}{\Gamma(4N)} \label{ratio}
\end{equation}
to good accuracy. We have "guessed" the peculiar factor in Eq.~(\ref{ratio}) by anticipating that the divergent series in Eq.~(\ref{series}) should yield the same tunneling factor as found in Ref.~\cite{Hatta:2007cs}.  The large $N$ part of the series in Eq.~(\ref{series}) can be written as
\begin{equation}
a\left(0, \rho \right) \sim \sum_{N}^{\infty} \Gamma(4N)\left(\frac{32\rho}{c^4}\right)^{N} \label{seriesN}
\end{equation}
or in terms of a Borel series
\begin{equation}
a\left(0, \rho \right) \sim \int_{0}^{\infty} \frac{\textrm{d}b}{b} \exp \left[-\frac{bc}{\left(32\rho\right)^{1/4}}\right]\sum_{N} \left(b^{4}\right)^N. \label{seriesB}
\end{equation}

\begin{center}
\begin{table}
\caption{Numerical evaluation of $a_{N}$. $C_N = \frac{a_{N+1}}{a_{N}} \frac{c^4\Gamma(4N)}{32\Gamma(4N+4)}$ with $c=\frac{2\Gamma^2\left(1/4\right)}{3\sqrt{2\pi}}$.}
  \begin{tabular}{ |l | c |c|c| c | }
    \hline
    $N$ & $a_{N}$ & $\frac{a_{N+1}}{a_{N}}$ & $\frac{32\Gamma(4N+4)}{c^4\Gamma(4N)}$& $C_N$\\ \hline
     4 & $1.7222\times 10^9$ & 20000 & 19926 & 1.0037\\ \hline
     5 & $3.4444\times 10^{13}$ & 45631 &45523 &1.0024 \\ \hline
     6 & $1.5717 \times 10^{18}$ & 90376 & 90222 &1.0017 \\ \hline
     7 & $1.4205 \times 10^{23} $ & $1.6196\times 10^{5}$&  $1.6176\times 10^5$& 1.0013\\ \hline
     8 & $2.3006 \times 10^{28}$ & $2.6944 \times 10^{5}$& $2.6918\times 10^5$& 1.0010 \\ \hline
     9 & $6.1987 \times 10^{33}$ &$4.2317\times 10^5$  & $4.2285\times 10^5$ & 1.0008  \\ \hline
     10& $2.6231 \times 10^{39}$ & $6.3483\times 10^5$ & $6.3444\times 10^5$ & 1.0006 \\ \hline
     11& $1.6652\times 10^{45}$ & $9.1743\times 10^5$ & $9.1695\times 10^5$ &1.0005 \\ \hline
     12 & $1.5277\times 10^{51}$ & $1.2853\times 10^6$ & $1.2847\times 10^6$ &1.0004 \\ \hline
     13 & $1.9635\times 10^{57}$ & $\cdots$ & $\cdots$& $\cdots$ \\
     \hline
  \end{tabular}
  \label{nume}
  \end{table}
\end{center}
Since $\sum_{N} \left(b^{4}\right)^N$ has a singularity at $b=1$, the $b$-integral is not well defined meaning the series Eq.~(\ref{series})  is not Borel summable. However, if we interpret Eq.~(\ref{series}) as an asymptotic series then Eq.~(\ref{seriesB}) suggests a non perturbative part, coming from $b=1$, which in general will not be purely real as are the individual terms in Eq.~(\ref{series}), of size
\begin{equation}
a\left(0, \rho \right) \sim \exp \left[-\frac{c}{\left(32\rho\right)^{1/4}}\right] \label{bt}
\end{equation}
which from Eq.~(\ref{ratio}) gives
\begin{equation}
a\left(0, \rho \right) \sim \exp \left[-c\sqrt{\frac{K^3}{k}}\right] \label{bt2}
\end{equation}
exactly the result of Ref.~\cite{Hatta:2007cs}.

\subsection{Finite length matter}
The multiple scattering series - operator product expansion for finite length matter is given by expanding the exponentials in Eq.~(\ref{a1m}) and Eq.~(\ref{h1m}). For small $\zeta$
\begin{equation}
a\left(\zeta\right)=a^{(0)}\left(\zeta\right)+\left[qA_{0}+\omega A_{3}\right]_{u=0}\frac{i q}{2\pi^2 T^2 x}\sum_{N=1}^{\infty} a_{N}\left(2i\rho LT x\right)^N \label{anf}
\end{equation}
with
\begin{equation}
a_{N} = \frac{1}{N !}\int_{0}^{\infty} \textrm{d} \zeta \zeta^{4N+1} \textrm{K}_{0}^{2}\left(\zeta\right) =  \frac{2^{4N-1}\Gamma^4(2N+1)}{\Gamma(N+1)\Gamma (4N+2)}.
\end{equation}
When $N$ is large
\begin{equation}
a_{N} \simeq \sqrt{\frac{\pi}{24N}}\left(\frac{16}{27}\right)^N \Gamma(3N+1).
\end{equation}
The sum in Eq.~(\ref{anf}) is divegent, but because of the $i^N$ factor it appears Borel summable and one can write
\begin{equation}
a\left(\zeta\right)=a^{(0)}\left(\zeta\right)+\frac{i q \left[qA_{0}+\omega A_{3}\right]_{u=0}}{2\pi^2 T^2 x}\int_{0}^{\infty}\frac{\textrm{d}b}{b}\exp \left[-\frac{b}{\left(2\rho LT x\right)^{1/3}}\right]\sum_{N=1}^{\infty} \frac{a_{N}\left(ib^3\right)^N}{\Gamma(3N)} \label{anfb}
\end{equation}
Now the $b$-integral is well defined and Eq.~(\ref{anfb}) and Eq.~(\ref{a1m}) are equivalent representations of $a\left(\zeta\right)$ for small $\zeta$.

\section{Dipole scattering on finite length hot matter}
\label{sec4}
A problem closely related to deep inelastic scattering on hot matter is the problem of the scattering of a high energy dipole on hot matter\cite{Albacete:2008ze}. Indeed in QCD these processes are closely related and at small-$x$ one can express deep inelastic scattering on a target in terms of dipole scattering on that target at least in the leading order renormalization group formalism. In strong coupling SYM theory the situation is not quite the same as in QCD. Deep inelastic scattering, in terms of $\mathcal{R}$-currents, involves only adjoint representation fields on the field theory side of the AdS/CFT correspondence while dipole scattering, given in terms of Wilson lines, involves fundamental representations external to the $\mathcal{N}=4$ SYM theory. This results in dipole scattering being intimately related to a string passing through the medium\cite{Peeters:2006iu, Liu:2006he, Chernicoff:2006hi, Caceres:2006ta, Maldacena:1998im, Rey:1998ik}. The rapidity at which cross sections become large, characterized by the saturation momentum, is the same in the two cases, as was the case for infinite matter, however, there appears to be no analog of the multiple scattering series - operator product expansion for dipole scattering, and thus structure functions defined in terms of dipole scattering are zero at $x$-values above the saturation value $x_{s}(Q)$, except for possible tunneling effects\cite{Faulkner:2008qk}.

As in earlier sections we imagine a strip of hot matter in the region $0\leq z \leq L$, $-\infty <x, y< \infty$ and we suppose a dipole impinges on the matter with the dipole moving in the $z$-direction from smaller values of $z$ to larger values. The ends of the dipole are connected by a string which we parametrize as
\begin{equation}
X^{\mu} = \left( t, x, 0, z, u\right)
\end{equation}
where we choose $x$ and $t$ to label points on the string so that
\begin{eqnarray}
z&=&z(t,x) \\
u&=&u(t,x)
\end{eqnarray}
and we now write the AdS black brane metric as
\begin{equation}
\textrm{d} s^2=R^2 u^2\left[-\left(1-\left(\frac{u_0}{u}\right)^4\right)\textrm{d} t^2+\textrm{d}
 \vec{x}^2+\frac{\textrm{d} u^2/u^4}{1-\left(\frac{u_0}{u}\right)^4}\right] \label{metric2}
\end{equation}
with $u_0 =\pi T$. The metric corresponding to the vacuum in the field theory is as in Eq.~(\ref{metric2}) but with $u_0$ set to zero. We choose the ends of the dipole to be at $\pm \frac{x_0}{2}$ so that the dipole "size" is $x_0$. We also take $u\left(t, \pm \frac{x_0}{2}\right) = \infty$ so the ends of the dipole are at the boundary. The Nambu-Goto action for the string in the medium is
\begin{equation}
S=-\frac{\sqrt{\lambda}}{2\pi R^2} \int \textrm{d}t \int_{-\frac{x_0}{2}}^{\frac{x_0}{2}} \textrm{d} x\sqrt{-g} = \int \textrm{d}t \int_{-\frac{x_0}{2}}^{\frac{x_0}{2}} \textrm{d}x\mathcal{L}
\end{equation}
where
\begin{equation}
-g=\left(uR\right)^4\left\{f(u)-\dot{z}^2-\frac{\dot{u}^2}{f(u)u^4}+z^{\prime 2}f(u)+\frac{u^{\prime 2}}{u^4}-\frac{\left(z^{\prime}\dot{u}-\dot{z} u^{\prime}\right)^2}{f(u)u^4}\right\} \label{g}
\end{equation}
with
\begin{equation}
f\left(u\right) = 1-\left(\frac{u_0}{u}\right)^4
\end{equation}
and where
\begin{equation}
\dot{u} =\frac{\partial u}{\partial t} \quad , \quad u^{\prime}=\frac{\partial u}{\partial x}
\end{equation}
and similarly for derivatives acting on $z$. One gets the metric, and string action, corresponding to the dipole outside the medium by setting $f$ to one in Eq.~(\ref{g}).

\subsection{Lowest order picture of dipole and string passing through the medium}
\label{lowest}
When the string approaches the strip of matter, we suppose that its shape and motion is simply the boost of its minimum energy static configuration so that $z^{\prime}=0$ and $\dot{z}(t,x)=v=\tanh \eta$ independent of $t$ and $x$. When $\cosh \eta \equiv \gamma$ is large, we expect the shape of the string to be unmodified as it passes through the medium, but we expect change in $\dot{u}$ and $\dot{z}$.

In Eq.~(\ref{g}) we set $z^{\prime}=0$ and find
\begin{equation}
\mathcal{L}=-\frac{\sqrt{\lambda}u^2}{2\pi \gamma}\left\{\left(1+\frac{u^{\prime 2}}{u^4-u_0^{4}}\right)\left(1-\frac{\gamma^2 u_0^4}{u^4}-\frac{\gamma^2 \dot{u}^2}{u^{\prime 2}+u^4-u_0^4}\right)\right\}^{1/2} \label{la}
\end{equation}
For the moment suppose $\frac{\gamma u_0^2}{u^2}$ and $\frac{\gamma \dot{u}}{u^2}$ are small so that the right hand side of Eq.~(\ref{la}) can be expanded to pick up corrections of size $\frac{\gamma^2 u_0^4}{u^4}$ but where corrections of size $\frac{u_0^4}{u^4}$ will be neglected. Then
\begin{equation}
\mathcal{L}\simeq-\frac{\sqrt{\lambda}}{2\pi \gamma}\sqrt{u^4+u^{\prime 2}}\left[1-\frac{\gamma^2 u_0^4}{2u^4}-\frac{\gamma^2 \dot{u}^2}{2\left(u^{\prime 2}+u^4\right)}\right]. \label{la1}
\end{equation}
Using this Lagrangian it is straightforward to find
\begin{equation}
\ddot{u}=-\frac{2u_0^4}{u} \label{u2}
\end{equation}
while the string goes through the medium. In obtaining Eq.~(\ref{u2}), we have dropped corrections proportional to $\dot{u}^2$ which, as we shall see shortly, are small compared to the terms that have been kept. Thus after passing through the medium the string has a changed motion given by
\begin{equation}
\Delta \dot{u} = -\frac{2 u_0^4 L}{u} . \label{udot}
\end{equation}
The change in the shape of the string after passing through the medium is given by
\begin{equation}
\Delta u = -\frac{u_0^4 L^2}{u}  \label{uchange}
\end{equation}
which, as expected, is higher order in $L$. We note that Eq.~(\ref{u2}) is identical to a result found in Ref.~\cite{Hatta:2008tx} where it appeared as $\frac{\textrm{d}\chi}{\textrm{d}\tilde{t}^2} = \frac{\chi^3}{8}$. Using $\chi = \frac{2 u_0}{u}$ and $\tilde{t}= 2 u_0 t$, we see that this is identical to Eq.~(\ref{u2}). The result of Ref.~\cite{Hatta:2007cs} was for the motion of a gravitational wave in the metric (\ref{metric2}) and it is interesting to see that over short times, where the genuine string properties coming from $u^{\prime}$ do not enter, a gravitational wave and a local part of the string are accelerated in an identical manner.

As the string passes from $z=0$ to $z=L$ , we have seen that $\dot{u}$  changes from $0$ to $-\frac{2 u_0^4 L}{u} $. However, the "potential" that the string passes through is time-independent so the energy of the string will not change. In addition, the time taken in passing from $0$ to $L$ is sufficiently short, when $L$ is not too large, so that there is little energy flow\cite{Herzog:2006gh, Gubser:2006bz} along the string and thus energy is approximately conserved locally on the string. Now the energy of the string outside of the interval $[0, L]$ is, assuming $z^{\prime}=0$,
\begin{equation}
E=\frac{\sqrt{\lambda}}{2\pi} \int_{-\frac{x_0}{2}}^{\frac{x_0}{2}} \textrm{d}x \frac{u^4}{u_m^2\left[1-\dot{z}^2-\frac{\dot{u}^2 u_m^4}{u^8}\right]^{1/2}} \label{energy}
\end{equation}
where we have used
\begin{equation}
E=\int_{-x_0 /2}^{x_0/2} \textrm{d}x \mathcal{H}
\end{equation}
and
\begin{equation}
\mathcal{H}=\dot{z}\frac{\partial \mathcal{L}}{\partial
\dot{z}}+\dot{u}\frac{\partial \mathcal{L}}{\partial
\dot{u}}-\mathcal{L}
\end{equation}
with $\mathcal{L}$ given in Eq.~(\ref{la}). We have also used
\begin{equation}
u^4+u^{\prime 2} = \frac{u^8}{u_m^4} \label{um}
\end{equation}
where $u_m$ is the minimum $u$-value on the string in the $AdS_5$ metric, when $f=1$. At $z=0$, $\dot{u}=0$, while at $z=L$, $\dot{u}$ is given by Eq.~(\ref{udot}). This requires a change in $\dot{z}$ in going from $z=0$ to $z=L$. Equating the energy locally in $u$, or in $x$, at $0$ and $L$ gives
\begin{equation}
-2\dot{z} \Delta \dot{z}= \frac{\left(\Delta \dot{u}\right)^2 u_m^4}{u^8}
\end{equation}
or, setting $\dot{z}\simeq 1$,
\begin{equation}
\Delta \dot{z}\simeq -2\frac{u_0^8 L^2 u_m^4}{u^{10}}. \label{dzv}
\end{equation}
We can now use this very small value of $\Delta \dot{z}$ to estimate $z^{\prime}$, which we have assumed to be zero up to now, as
\begin{equation}
\frac{dz}{dx} =\frac{dz}{du} \frac{du}{dx} \simeq\frac{d(\Delta \dot{z}L)}{du} \frac{du}{dx}\propto \frac{\Delta \dot{z}}{u} L u^{\prime}.
\end{equation}
Thus $z^{\prime}$ is of order $L^3$, much smaller than terms with which we have been interested in this section.

After the dipole, and string, pass through the medium they are back in the metric given by Eq.~(\ref{g}), but with $f=1$. The motion of the string then follows the equations determined by the Lagrangian $\mathcal{L}=-\frac{\sqrt{\lambda}}{2\pi} \sqrt{-g}$ with $g$ given by Eq.~(\ref{g}) and with the initial conditions given by taking $u_0(x)$ to be that of a free boosted string along with initial velocities set by Eq.~(\ref{udot}) and $\dot{z}_0(x)=v+\Delta \dot{z}\left(u(x)\right)$ where $v$ is the velocity of the string before passing through the medium. We are not here going to attempt to describe the motion of the string after leaving the medium, however, we do wish to determine at what values of the parameters $\gamma$ and $L$ the string has been sufficiently disturbed by the medium that it "splits". (We use the word "split" in the following sense. As we argue below from the field theory side of the AdS/CFT correspondence radiation is expected to occur when fields lag too far behind the ends of the dipole which are their sources. By the infrared-ultraviolet correspondence\cite{Susskind:1998dq, Peet:1998wn, Chesler:2007sv} we can relate the position of the energy of fields to the position of the string. Thus when a part of the string lags too far behind, in the $z$-direction, the ends of the dipole that part of the string should correspond to radiation and should separate from the string which at late times connects the ends of the dipoles. It is a challenging problem to understand how the string evolves in order to represent both the free dipole and the produced radiation at late times. However, it is clear that this requires string loop effects which we are not including here. At the level that oscillations on the string are quantized, a strong disturbance of the string will create excited states of the string. So long as loop corrections are absent these excitations are stable. However, even a very small string coupling, $g_s$, will allow these excitations to (ultimately) decay. For this reason we feel it is reasonable to describe the discussion below as string splitting even though the actual splitting will not occur until string loops are added.) It is fairly easy to do an \underline{estimate} when this happens, although we do not know how to do a precise calculation. It is easiest to "split" the string near its lowest point in $u$, $u_m$, which occurs at $x=0$. To understand when the string must break it is convenient to view the dipole on the field theory side of the AdS/CFT correspondence where one has fields connecting the dipole rather than a string. By the infrared-ultraviolet correspondence, the extent of the fields is $\frac{1}{u(x)}$ in the $x$ and $y$ directions and
\begin{equation}
\left(\Delta z\right)_d \sim \frac{1}{\gamma u(x)}, \label{dz}
\end{equation}
at $u$, in the $z$-direction. The $1/\gamma$ factor in Eq.~(\ref{dz}) is the Lorentz contraction factor. Still on the field theory side, the lifetime of quanta at $u$ is $\gamma/u$. Thus when the dipole emerges from the matter, the change in velocity given in Eq.~(\ref{dzv}) persists over a time $\gamma/u$ before interactions can modify Eq.~(\ref{dzv}) . If, during this time the fields fall too far behind the ends of the dipole, they should correspond to radiation rather than part of the dipole. Thus the string will split if
\begin{equation}
\Delta \dot{z} \frac{\gamma}{u} > \frac{1}{\gamma u}.
\end{equation}
Using Eq.~(\ref{dzv}) string splitting requires
\begin{equation}
\frac{u_0^4 L u_m^2 \gamma }{u^5}> 1
\end{equation}
which at $u=u_m$ becomes
\begin{equation}
\frac{u_0^4 L\gamma }{u_m^3}> 1. \label{opef}
\end{equation}
Recalling that $\frac{u_0 \gamma}{u} \sim \frac{1}{x} $, Eq.~(\ref{opef}) is the same as Eq.~(\ref{sam}). Thus for finite length matter complete absorption of $\mathcal{R}$-currents occurs at the same values of $\gamma$ and $L$ as for string splitting when a dipole passes through the matter. This is similar to the case of infinite extent matter where it was noted that string breaking and a non zero $F_2$ structure function for $\mathcal{R}$-currents occur, parametrically, at the same rapidity.

Finally, we come to a subtle point in our discussion. We have argued that string splitting requires $\frac{u_0^4 L\gamma }{u_m^3}> 1$. But at even smaller values of $\gamma$, the Lagrangian given by Eq.~(\ref{la}) for the string in the medium becomes complex, a phenomenon which seems to occur at
\begin{equation}
\gamma \frac{u_0^2}{u_m^2}\simeq 1 \label{gm}
\end{equation}
when $\gamma$ gets as large as the value indicated in Eq.~(\ref{gm}), the expansion going from  Eq.~(\ref{la}) to Eq.~(\ref{la1}) ceases to be valid. We are now going to give a more complete mathematical analysis using Eq.~(\ref{g}) when $\gamma$ is large. However, as we shall see, the results of the lowest order expansion given in Eq.~(\ref{la1}) and leading to Eqs.~(\ref{udot}), (\ref{dzv}) and (\ref{opef}) are in fact valid even when $\gamma$ is larger the value determined by Eq.~(\ref{gm}).

\subsection{More complete picture of dipole and string passing through the medium}
\label{more}
Now we use the more complete metric Eq.~(\ref{g}) and the Lagrangian $\mathcal{L}=-\frac{\sqrt{\lambda}}{2\pi R^2}\sqrt{-g}$ coming from that metric. We suppose the dipole, and string, approaches the medium with $\gamma$ where $\gamma \frac{u_0^2}{u_m^2}\gg 1$. As the string enters the medium it must slow down to conserve energy. We are working in an approximation where the slowing down is instantaneous as the string enters the medium. Just before entering the medium the energy is given by Eq.~(\ref{energy}) with $\dot{u}$ set to zero. After entering the medium the energy density is given by
\begin{equation}
 \mathcal{H}=\frac{\sqrt{\lambda}}{2\pi} \frac{1+\frac{u^{\prime 2}}{u^4}}{\left\{\quad\right\}^{1/2}}u^2 \label{energy2}
\end{equation}
with
\begin{equation}
\left\{\quad\right\}=\left(1-\dot{z}^2-\frac{u_0^4}{u^4}\right)\left(1+\frac{u^{\prime 2}}{u^4-u^4_0}\right)-\frac{\dot{u}^2}{u^4-u^4_0}+z^{\prime 2}\left(1-\frac{u^4_0}{u^4}\right)+\frac{2z^{\prime} \dot{z} \dot{u} u^{\prime}}{u^4-u^4_0}, \label{den}
\end{equation}
where $z=z(t, x)$. As we shall see below $z^{\prime}$ is now of the same size as $\frac{\dot{u}}{u^2}$ and so $z^{\prime}$-terms in Eq.~(\ref{energy2}) must be kept, however, we have dropped $z^{\prime 2}$ terms in the numerator of Eq.~(\ref{energy2}) and we have dropped $\left(\dot{u} z^{\prime}\right)^2$ terms in Eq.~(\ref{den}). Equating Eq.~(\ref{energy2}) with the integrand of Eq.~(\ref{energy}) gives
\begin{equation}
1-\dot{z}^2 -\frac{u^4_0}{u^4} \simeq 1/\gamma^2 \quad \text{or} \quad \dot{z} \simeq 1-\frac{1}{2} \frac{u^4_0}{u^4} \label{zdot}
\end{equation}
as the string passes through the medium. $\gamma$ in Eq.~(\ref{zdot}) is given by $1/\gamma^2=1-v^2$ where $v$ is the velocity \underline{before} entering the medium, the $\dot{z}$ appearing in Eq.~(\ref{energy}), but of course distinct from the $\dot{z}$ in Eq.~(\ref{zdot}) which is the velocity \underline{in} the medium. The equations of motion are
\begin{equation}
\frac{\textrm{d}}{\textrm{d} t} \frac{\partial \mathcal{L}}{\partial \dot{u}} + \frac{\textrm{d}}{\textrm{d} x} \frac{\partial \mathcal{L}}{\partial u^{\prime}}-\frac{\partial \mathcal{L}}{\partial u}=0. \label{eom}
\end{equation}
It is straightforward to find
\begin{equation}
\frac{\partial \mathcal{L}}{\partial \dot{u}} \simeq \frac{\sqrt{\lambda}}{2\pi } \frac{u^2}{\left\{\quad\right\}}\frac{\dot{u}-u^{\prime} z^{\prime} }{u^4} \label{ludot}
\end{equation}
where we have neglected factor of $\frac{u^4_0}{u^4}$ and terms of size $\dot{u} z^{\prime}$ compared to $\dot{u}$. We have also set $\dot{z}=1$ in the numerator to get Eq.~(\ref{ludot}). Similarly
\begin{equation}
\frac{\partial \mathcal{L}}{\partial u} \simeq -\frac{\sqrt{\lambda}}{2\pi } \frac{u^2}{\left\{\quad\right\}}\frac{2 u^4_0 }{u^5} \left(1+\frac{u^{\prime 2}}{u^4}\right) \label{lu}
\end{equation}
and
\begin{equation}
\frac{\partial \mathcal{L}}{\partial u^{\prime}} \simeq -\frac{\sqrt{\lambda}}{2\pi } \frac{u^2}{\left\{\quad\right\}}\left(1-\dot{z}^2-\frac{u^{4}_0}{u^4}\right)\frac{ u^{\prime} }{u^4} \label{luprime}
\end{equation}
Using Eq.~(\ref{zdot}) we see that Eq.~(\ref{luprime}) is very small so using Eqs.~(\ref{ludot}) and (\ref{lu}) in Eq.~(\ref{eom}) gives
\begin{equation}
\ddot{u}-u^{\prime} \dot{z}^{\prime} \simeq -\frac{2u_0^4}{u} \left(1+\frac{u^{\prime 2}}{u^4}\right)
\end{equation}
which, after again using Eq.~(\ref{zdot}) gives
\begin{equation}
\ddot{u}=-\frac{2u_0^4}{u} \nonumber
\end{equation}
 exactly as in Eq.~(\ref{u2}).

Now when the string leaves the medium, it again gains velocity so as to conserve energy so that $\Delta \dot{z}$ becomes
\begin{equation}
\Delta \dot{z} =-\frac{2u^8_0 L^2}{u^6} \label{changezdot}
\end{equation} 
which replaces Eq.~(\ref{dzv}) and agrees with that equation near the lowest point on the string. The only essential difference between our present discussion and our lowest order discussion is in the value of $\Delta z$. Earlier we found, from Eq.~(\ref{dzv}), that
\begin{equation}
\Delta z = -\frac{u_0^8 L^3 u_m^4}{u^{10}} \label{zchange2}
\end{equation}
just after emerging from the medium. Now Eq.~(\ref{zdot}) gives
\begin{equation}
\Delta \overline{ z} =-\frac{1}{2} \frac{u_0^4}{u^4} L \label{zbchange}
\end{equation}
where we use $\overline{z}$ rather than $z$ simply to distinguish the $\Delta \overline{ z} $ in Eq.~(\ref{zbchange}) from the lowest order calculation given in Eq.~(\ref{zchange2}) and based on Eq.~(\ref{la1}). $\Delta \overline{ z} $ is greater than $(\Delta z)_d$ given in Eq.~(\ref{dz}) when $\frac{\gamma u_0^4 L}{u^3} >1$ exactly as in Eq.~(\ref{opef}), but now it appears that the string breaking is determined immediately after leaving the medium while earlier we had to follow the evolution of the string over a time $\gamma/u_m$ to see a separation, in $z$, large enough that it corresponded to string splitting. However, since the coherence time of the string at $u_m$ is $\frac{\gamma}{u_m}$, it is not clear that the difference between Eq.~(\ref{zchange2}) and Eq.~(\ref{zbchange}) is significant. We note that if we were to consider a Lorentz frame where the strip of matter were located in a region
\begin{equation}
0 \leq  z+t\leq \frac{L}{\cosh \eta_1},
\end{equation}
with $\eta_1$ the boost from our frame where the matter is at rest , then the metric would look like that of a shock wave\cite{Albacete:2008ze, Janik:2005zt, Kancheli:2002nw} and the $\Delta \overline{z}$ of Eq.~(\ref{zbchange}) would become a time delay during which the part of the string at $u$ is captured by the shock wave. The capture time\cite{Kancheli:2002nw} in this frame is
\begin{equation}
\left(\Delta t\right)_{\text{capture}} = -\Delta \overline{z} \cosh \eta_{1}, \label{traptime}
\end{equation}
where $\Delta \overline{z}$ is given in Eq.~(\ref{zbchange}).

Indeed, it is rather easy to give a heuristic derivation of Eqs.~(\ref{zbchange}) and (\ref{traptime}) as well as (\ref{u2}). Using the shock wave metric \cite{Albacete:2008ze, Janik:2005zt}
\begin{equation}
\textrm{d} s^2=R^2 u^2\left[-2 \textrm{d}x^{+} \textrm{d}x^{-}+ \sqrt{2}L \cosh \eta_1\left(\frac{u_0}{u}\right)^4 \delta\left(x^{+}\right)\textrm{d} x^{+2}+\textrm{d}
 x^2_{\perp}+\frac{\textrm{d} u^2}{u^4}\right] \label{metric3}
\end{equation}
where $x^{\pm}=\frac{1}{\sqrt{2}}\left(t\pm z\right)$ and where $\eta_1$ is the rapidity of the shock wave and $\eta -\eta_1$ is the rapidity of the dipole (and the string) as it approaches the shock wave. As we have seen in our previous calculations, and this is even clearer in the present circumstance, the different parts of the string act independently in passing through the matter. Therefore, the motion of the part of the string at $u(x,t)$ should be determined by requiring that element of the string lie on a null geodesic as it pass through the shock wave\cite{Hatta:2008tx, Sin:2004yx}. Writing Eq.~(\ref{metric3}) as
\begin{equation}
\textrm{d} s^2=-2R^2 u^2 \textrm{d}x^{+} \left[\textrm{d}x^{-}- \frac{L \cosh \eta_1}{\sqrt{2}} \left(\frac{u_0}{u}\right)^4 \delta\left(x^{+}\right)\textrm{d} x^{+}-\frac{1}{2u^4}\left(\frac{\textrm{d} u}{\textrm{d}x^{+}}\right)^2\textrm{d}x^{+}\right],\label{metric4}
\end{equation}
where we have dropped the $\textrm{d}x_{\perp}^2$ term which plays no role in our discussion, we see immediately that being on a null geodesic gives a jump in $\Delta x^{-}$, at $x^{+}=0$, of size
\begin{equation}
\Delta x^{-}=\frac{L \cosh \eta_1}{\sqrt{2}} \left(\frac{u_0}{u}\right)^4 \label{null}
\end{equation}
as the string passes through the shock wave. This is exactly the same as Eq.~(\ref{traptime}). Thus as the string emerges from the shock wave $z(u)$ has a nontrivial $u$-dependence given by
\begin{equation}
z\left(u, t\right)-z\left(\infty, t\right)= -\frac{1}{2} \left(\frac{u_0}{u}\right)^4 L \label{shockz}
\end{equation}
where Eq.~(\ref{shockz}) is written in the rest frame of the medium. But a $u$-dependence of $z$ requires an energy flow along the string given by Ref.~\cite{Herzog:2006gh}, in the rest frame of the medium given by
\begin{equation}
\frac{\textrm{d} E}{\textrm{d} t} = \pi_{t}^{1}=\frac{\sqrt{\lambda}}{2 \pi } u^4 \cosh \eta \frac{\textrm{d} z}{\textrm{d} t} \frac{\textrm{d}z}{\textrm{d} u}. \label{flow1}
\end{equation}
Recalling that $\frac{\sqrt{\lambda}}{2\pi} \cosh \eta$ is just the energy density of the string, the energy flow must also be given by
\begin{equation}
\frac{\textrm{d} E}{\textrm{d} t} =\frac{\sqrt{\lambda}}{2 \pi } \cosh \eta \frac{\textrm{d} u}{\textrm{d} t}. \label{flow2}
\end{equation}
Thus, equating Eqs~(\ref{flow1}) and (\ref{flow2}), and using (\ref{shockz})  and $\frac{\textrm{d} z}{\textrm{d} t} \simeq 1$, gives
\begin{equation}
\Delta \dot{u} = - \frac{2 u_0^4}{u}L \nonumber
\end{equation}
the same as we found earlier in Eqs~(\ref{u2}) and (\ref{udot}).

Our present calculation is mathematically much more realiable than the calculation we give in Sec.~\ref{lowest}, however, that calculation had the intuitive appeal of the motion of the string in the medium being completely determined by single graviton exchange acting on an otherwise free string passing through the region of the medium. It would be interesting to better understand the relationship between these two procedures of calculation.

\appendix
\section{The double scattering contribution}
\label{secondorder}
Here we evaluate the double scattering contribution to Eqs.~(\ref{a1m}) and (\ref{h1m}) and, using Eq.~(\ref{action1}), determine $F_1$ and $F_L$ at this leading higher twist level. Expanding the exponential in Eqs.~(\ref{a1m}) and (\ref{h1m}) to second order in $\tilde{L}$  and evaluating the integrals, one finds,
\begin{equation}
a^{(2)}\left(\zeta\right) = -i q^2 A_{L} (0)\frac{(2\pi)^{6}}{35} \frac{L^2 T^4}{Q^4}\left(\frac{1}{x}\right)^3 \label{a2}
\end{equation}
and
\begin{equation}
h^{(2)}\left(\zeta\right) = i \zeta A_{i} (0)\frac{(2\pi)^{8}}{112} \frac{L^2 T^6}{Q^4}\left(\frac{1}{x}\right)^3 \label{h2}
\end{equation}
for small $\zeta$. Using Eqs~(\ref{a2}) and (\ref{h2}) in Eq.~(\ref{action1}), with the help of Eqs.~(\ref{al}) and (\ref{ai2}), one finds
\begin{equation}
S=\frac{iN^2\pi^6}{7x} \left(\frac{LT^3}{xQ^2}\right)^2 \left[\frac{4}{5}(qA_{0}+\omega A_{3})^2+Q^2 A_{i}^2\right]_{u=0}.
\end{equation}
Using Eqs.~(\ref{f1d}), (\ref{f2}), (\ref{r12}) and (\ref{SR}), one gets
\begin{equation}
F_1 = \frac{N^2 Q^2\pi ^5}{7x} \left(\frac{LT^3}{xQ^2}\right)^2
\end{equation}
and
\begin{equation}
F_L = F_2- 2x F_1= \frac{8}{5} \frac{N^2 Q^2\pi ^5}{7} \left(\frac{LT^3}{xQ^2}\right)^2.
\end{equation}
The above equations are reliable when $\frac{LT^3}{xQ^2} \ll 1$.

\section{Dipole scattering on finite hot medium in the rest frame of the dipole}
\label{restframe}
Here we carry out the calculation of dipole scattering on finite hot medium in the dipole rest frame. Let us boost the metric in Eq.~(\ref{metric2}) to the rest frame of the dipole by using
\begin{eqnarray}
\textrm{d}t &=& \gamma \textrm{d}t^{\prime} +\gamma v \textrm{d}z^{\prime}, \\
\textrm{d}z &=& \gamma v \textrm{d} t^{\prime} +\gamma \textrm{d} z^{\prime}.
\end{eqnarray}
After dropping the primes, it is easy to find the transformed metric
\begin{equation}
\textrm{d} s^2=R^2 u^2\left[-\left(1-\frac{\gamma^2 u_0^4}{u^4}\right)\textrm{d} t^2+\textrm{d}x^2_{\perp}+\left(1+ \frac{\gamma^2 v^2 u_0^4}{u^4}\right)\textrm{d} z^2+2\textrm{d}t \textrm{d}z \frac{\gamma^2 v u_0^4}{u^4} +\frac{\textrm{d} u^2/u^4}{1-\frac{u_0^4}{u^4}}\right]. \label{metricshock}
\end{equation}
Eq.~(\ref{metricshock}) is essentially the same as the shock wave metric (Eq.~(\ref{metric3})) with $\gamma = \cosh \eta_1$, and is equivalent to Eq. (3.7) of Ref.~\cite{Albacete:2008ze} in the large $\gamma$ limit.

If one puts the dipole at rest at $z=0$ and assumes that the medium has a width of $L$ at its rest frame,  one easily sees that the Lagrangian $\mathcal{L}$ actually is time-dependent. At $\tau_0=0$, the medium starts to interact with the dipole, while at $\tau_1 = \frac{L}{\gamma}$ the medium passes by the dipole and the interaction stops. Therefore, the energy of the string is no longer conserved, but the $+$-component of momentum is now a conserved quantity. (We recall that the energy of the string is conserved in the medium rest frame, while the $z$-component of momentum is not.)
In this frame, one finds
\begin{equation}
\ddot{u}=-\frac{2\gamma^2 u^4_0}{u} \label{u2r},
\end{equation}
and
\begin{equation}
\Delta \dot{u}=-\frac{2\gamma u^4_0 L}{u} \label{u1r}.
\end{equation}
Noting that $\gamma \textrm{d} \tau =\textrm{d} t$, it is easy to see that Eqs.~(\ref{u2r}) and (\ref{u1r}) are essentially equivalent to Eqs~(\ref{u2}) and (\ref{udot}) as we found above. After the medium passes by the dipole when $\tau > \frac{L}{\gamma}$, the Lagrangian and the energy density are given by
\begin{equation}
\mathcal{L} = -\frac{\sqrt{\lambda}}{2\pi}\sqrt{\left(u^4+u^{\prime 2}\right)-\Delta\dot{u}^2+ u^4 \Delta z^{\prime 2}-u^4 \Delta \dot{z}^2- \left(u^{\prime} \Delta \dot{z}-\Delta\dot{u} \Delta z^{\prime}\right)^2},
\end{equation}
and
\begin{eqnarray}
\mathcal{H} &= &\frac{\sqrt{\lambda}}{2 \pi} \frac{u^4+u^{\prime 2} +u^4\Delta z^{\prime 2}}{\sqrt{u^4+u^{\prime 2} - \Delta\dot{u}^2+ u^4\Delta z^{\prime 2}-u^4 \Delta \dot{z}^2- \left(u^{\prime} \Delta \dot{z}-\Delta \dot{u} \Delta z^{\prime}\right)^2 } }\\ \label{hr0}
  &\simeq& \frac{\sqrt{\lambda}}{2 \pi} \sqrt{u^4+u^{\prime 2}}\left[1+\underbrace{\frac{1}{2}\frac{\Delta \dot{u}^2}{u^4+u^{\prime 2}}}_{\textrm{Term A}}+\underbrace{\frac{1}{2}\frac{u^4 \Delta z^{\prime 2}}{u^4+u^{\prime 2}}}_{\textrm{Term B}}+\underbrace{\frac{1}{2}\frac{u^4\Delta \dot{z}^2}{u^4+u^{\prime 2}}}_{\textrm{Term C}}+\underbrace{\frac{\left(u^{\prime} \Delta \dot{z}-\Delta\dot{u} \Delta z^{\prime}\right)^2}{2\left(u^4+u^{\prime 2}\right)}}_{\textrm{Term D}}\right], \label{hr}
\end{eqnarray}
respectively. For the moment we assume that \textrm{Term A-D} are small in order to justify the Taylor expansion (see later). Using Eq.~(\ref{um}) and the first term in Eq.~(\ref{hr}), integrating over the whole string and subtracting the bare mass of the two quarks, one can find the potential energy between the quark-antiquark pair
\begin{equation}
\textrm{V}=\textrm{E}-2m_0 = -\kappa \frac{\sqrt{\lambda}}{\pi} u_m, \label{potential}
\end{equation}
where $\kappa = \frac{\sqrt{\pi} \Gamma (3/4)}{\Gamma(1/4)}$ and $u_m$ is the minimum position of the string. It is straightforward to relate $u_m$ and $x_0$(the size of the dipole) and obtain $x_0 = \frac{2\kappa}{u_m}$, thus $\textrm{V} = -\frac{\sqrt{\lambda}}{2\pi} \frac{4\kappa^2}{x_0} $.

Before we start to evaluate the following corrections in the square bracket, we recall the results obtained in Eqs.~(\ref{changezdot}) and (\ref{zbchange}), transform them into the rest frame of the dipole and get 
\begin{eqnarray}
\Delta \dot{z} &=& -\gamma^2 \frac{2u^8_0 L^2}{u^6}  \label{rdz}\\
\Delta z^{\prime} &=&\gamma \frac{2u_0^4 L u^{\prime}}{u^5}  . \label{rpz}
\end{eqnarray}

Integrating over the $\textrm{Term A}$ together with the factor outside of the square bracket gives the first contribution to the kinetic energy ($\textrm{KE}_A$) of the string
\begin{equation}
\textrm{KE}_A = \int_{-x_0/2}^{x_0/2} \textrm{d}x \frac{\sqrt{\lambda}}{2 \pi} \frac{1}{2}\frac{\dot{u}^2}{\sqrt{u^4+u^{\prime 2}}}=\frac{\sqrt{\lambda}}{2 \pi} \frac{5\pi}{21 \kappa} \frac{\gamma^2 u_0^8 L^2}{u_m^5}. \label{kinetic}
\end{equation}
Evaluating $\textrm{Term B}$ yields the second contribution to the kinetic energy ($\textrm{KE}_B$) of the string
\begin{equation}
\textrm{KE}_B = \int_{-x_0/2}^{x_0/2} \textrm{d}x \frac{\sqrt{\lambda}}{2 \pi} \frac{1}{2}\frac{u^4 \Delta z^{\prime 2}}{\sqrt{u^4+u^{\prime 2}}}=\frac{\sqrt{\lambda}}{2 \pi} \frac{2\pi}{21 \kappa} \frac{\gamma^2 u_0^8 L^2}{u_m^5}. \label{kinetic2}
\end{equation}

Comparing Eqs~(\ref{kinetic}) and (\ref{kinetic2}) with Eq.~(\ref{potential}), it is easy to see that the expansion parameter that we have assumed in Eq.~(\ref{hr}) is $\frac{\gamma u^4_0 L}{u_m^3}$. This parameter is essentially the same as the OPE parameter that we used in Eq.~(\ref{anf}). Furthermore, it is straightforward to estimate the contributions from $\textrm{Term C}$ and $\textrm{Term D}$ using Eqs~(\ref{rdz}) and (\ref{rpz}), and find that 
\begin{eqnarray}
\textrm{KE}_C &\sim& \frac{\sqrt{\lambda}}{2 \pi} u_m \left(\frac{\gamma u^4_0 L}{u_m^3}\right)^4 \\ 
\textrm{KE}_D &\sim& \frac{\sqrt{\lambda}}{2 \pi} u_m \left(\frac{\gamma u^4_0 L}{u_m^3}\right)^4.
\end{eqnarray}
Here we do not evaluate the coefficients of the contributions from $\textrm{Term C}$ and $\textrm{Term D}$, because they are of higher order in terms of the expansion parameter $\frac{\gamma u^4_0 L}{u_m^3}$. Therefore, one gets the total kinetic energy $\textrm{KE}$ by summing from $\textrm{Term A}$ to $\textrm{Term D}$ as well as higher order terms in the expansion
\begin{eqnarray}
\textrm{KE} &=& \textrm{KE}_A + \textrm{KE}_B+\textrm{KE}_C+\textrm{KE}_D+\cdots\\
  &\simeq& \frac{\sqrt{\lambda}}{2 \pi} \frac{\pi}{3 \kappa} \frac{\gamma^2 u_0^8 L^2}{u_m^5}\left[1+\mathcal{O} \left(\frac{\gamma u^4_0 L}{u_m^3}\right)^2\right]
\end{eqnarray}

When
\begin{equation}
\textrm{V}+ \textrm{KE} \geq 0, \label{total}
\end{equation}
the dipole starts to dissociate. Thus, this condition corresponds
to $\frac{\gamma u_0^4 L}{u_m^3} > \mathcal{O}\left(1\right)$,
which is in agreement with Eq.~(\ref{opef}) and Eq.~(\ref{sam}).
Of course, when Eq.~(\ref{total}) is satisfied the expansion leading from Eq.~(\ref{hr0}) to Eq.~(\ref{hr}) breaks down. Nevertheless, we expect Eq~(\ref{total}), using Eq~(\ref{potential}) and Eq~(\ref{kinetic}) to be a reasonable estimate of when string splitting should become important. Eventually, the kinetic energy ($\textrm{KE}$) should turn into
radiation at very late time. In the end, we also note that there exists a limiting length
\begin{equation}
x_L \simeq 1.1\frac{1}{\left(\gamma u_0^4 L\right)^{1/3}} \sim \frac{1}{\pi T \left(\gamma \pi T L\right)^{1/3}}
\end{equation}
for a finite length plasma according to Eq~(\ref{total}). There should be strong radiation and dipole dissociation when the size of the incoming dipole $x_0$ is larger than this limiting length $x_L$. 

\begin{acknowledgments}
A.Sh. wishes to thank A.H. Mueller and B. Xiao  for hospitality
and support during his visit at Columbia University in August 2008
when this  work was initiated.  A. Sh. acknowledges  financial
support by the Deutsche Forschungsgemeinschaft under contract
Sh 92/2-1. This work was supported in part by the US Department of Energy. 
\end{acknowledgments}

\end{document}